\documentclass{llncs}
\usepackage{llncsdoc}
\usepackage{cite}      
\usepackage{graphicx}  
\usepackage{subfigure} 
\usepackage{amsmath}   

\begin{document}
\title{Revisiting Internet AS-level Topology Discovery}

\author{ Xenofontas A. Dimitropoulos\inst{1} \and Dmitri V. Krioukov\inst{2} 
	\and George F. Riley\inst{1}}
\institute{School of Electrical and Computer Engineering\\
	Georgia Institute of Technology\\
	Atlanta, Georgia 30332--0250\\
	\email{fontas@ece.gatech.edu}\\
	\email{riley@ece.gatech.edu}
\and
  Cooperative Association for Internet Data Analysis (CAIDA)\\
  La Jolla, California 92093--0505\\
  \email{dima@caida.org}
}

\newcommand{\goodgap}{%
  \hspace{\subfigtopskip}%
  \hspace{\subfigbottomskip}}

\maketitle

\begin{abstract}
  The development of veracious models of the Internet topology has received
  a lot of attention in the last few years. Many proposed models are
  based on topologies derived from RouteViews~\cite{routeviews} BGP table
  dumps (BTDs). However, BTDs do not capture all AS--links of the Internet
  topology and most importantly the number of the hidden AS--links is
  unknown, resulting in AS--graphs of questionable quality.
  As a first step to address this problem, we introduce a new AS--topology
  discovery methodology that results in more complete and accurate graphs.
  Moreover, we use data available from existing measurement facilities,
  circumventing the burden of additional measurement infrastructure.
  We deploy our methodology and construct an AS--topology that has at
  least 61.5\% more AS--links than BTD--derived AS--topologies we examined.
  Finally, we analyze the temporal and topological properties of the
  augmented graph and pinpoint the differences from BTD--derived AS--topologies.
\end{abstract}

\section{Introduction}
Knowledge of the Internet topology is not merely of technological interest, 
but also of economical, governmental, and even social concern. As a result, 
discovery techniques have attracted substantial attention in the last few years. 
Discovery of the Internet topology involves passive or active measurements to 
convey information regarding the network infrastructure. We can use topology 
abstraction to classify topology discovery techniques into the following three 
categories: AS--, IP-- and LAN--level topology measurements. 
In the last category, SNMP--based as well as active probing techniques 
construct moderate size networks of bridges and end-hosts. At the IP--level (or 
router--level), which has received most of the research interest, discovery 
techniques rely on path probing to assemble WAN router--level 
maps~\cite{govindan00heuristics,skitter,sigcomm2002-rocketfuel}. Here, the 
two main challenges are the resolution of IP aliases and the sparse coverage 
of the Internet topology due to the small number of vantage points. 
While the latter can be ameliorated by increasing the number of measurement points 
using overlay networks and distributed agents~\cite{Scriptroute,DIMES,TR_AT_HOME},
the former remains a daunting endeavor addressed only partially thus 
far~\cite{SpDoRoWe04,iffinder}.
AS--level topology discovery has been the most straightforward, since 
BGP routing tables, which are publicly available in RouteViews (RV)~\cite{routeviews}, 
RIPE~\cite{RIPE} and several other Route Servers~\cite{traceroute-org}, 
expose parts of the Internet AS--map. However, the 
discovery of the AS--level topology is not as simple as it appears.

The use of BTDs to derive the Internet AS--level topology is a common method.
Characteristically, the seminal work by Faloutsos {\em et al.}~\cite{FFF99} 
discovered a set of simple power--law relationships that govern AS--level 
topologies derived from BTDs.
Several followup works on topology modeling, evolution modeling and synthetic topology 
generators have been based on these simple power--law properties~\cite{AiChLu00,
WilRevisited,tangmunarunkit02network}. However, it is well--known among the research 
community that the accuracy of BTD--derived topologies is arguable. First, a BGP 
table contains a list of AS--paths to destination prefixes, which do not necessarily 
unveil all the links between the ASs. For example, assume that the Internet topology 
is a hypothetical full mesh of size $n$, then from a single vantage point,
the shortest paths to every destination would only reveal $n-1$ of the 
total $n(n-1)/2$ links. In addition, BGP policies limit the export and import 
of routes. In particular, prefixes learned over peering 
links\footnote{``Peering links'' refers to the AS--relationship, in which 
two ASs mutually exchange their customers' prefixes free of charge.}
do not propagate upwards in the customer-provider hierarchy. Consequently, higher
tier ASs do not see peering links between ASs of lower tiers. This is one reason
BTD--based AS--relationships inference heuristics~\cite{gao00inferring} find only
a few thousands of peering links, while the Internet Routing Registries reveal tens of
thousands~\cite{SiFa04}. Lastly, as analyzed comprehensively in~\cite{TeRe04}, RV
servers only receive partial views from its neighboring routers, since the eBGP 
sessions filter out backup routes. 

The accuracy of AS--level topologies has been considered previously. 
In~\cite{chang04towards} Chang {\em et al.} explore several diverse 
data sources, i.e. multiple BTDs, Looking Glass servers 
and Internet Routing Registry (IRR) databases, to create a 
more thorough AS--level topology.
They report 40\% more connections than a BTD-derived AS--map and find 
that the lack of connectivity information increases for smaller degree ASs. 
Mao {\em et al.}~\cite{MaReWaKa03} develop a methodology to map router--graphs 
to AS--graphs. However they are more concerned with the methodology rather 
then the properties of the resulting AS--graph. Finally, in~\cite{AnFeBaBa02}
Andersen {\em et al.} explore temporal properties of BGP updates to create a 
correlation graph of IP prefixes and identify clusters. The clusters imply 
some topological proximity, however their study is not concerned with the 
AS--level topology, but rather with the correlation graph. 

Our methodology is based on exploiting BGP dynamics to discover additional 
topological information.
In particular we accumulate the AS--path information from BGP updates seen from RV
to create a comprehensive AS--level topology. The strength of our approach relies on 
a beneficial side--effect of the problematic nature of BGP convergence process. 
In the event of a routing change, the so-called ``path exploration'' problem, 
\cite{labovitz98internet}, results in superfluous 
BGP updates, which advertise distinct backup AS--paths of increasing length. 
Labovitz {\em et al.}~\cite{labovitz98internet} showed that there can be up to $O(n!)$ 
superfluous updates during BGP convergence.  We analyze these updates and find 
that they uncover a substantial number of new AS--links not seen previously. 
To illustrate this process, consider the simple update sequence 
in Table~\ref{example}, which was found in our dataset. The updates are 
received from a RV neighbor in AS10876 and pertain to the same 
prefix. The neighbor initially sends a withdrawal for the prefix 
205.162.1/24, shortly after an update for the same prefix that exposes 
the unknown to that point AS--link 2828--14815, and 
finally an update for a shorter AS--path, in which it converges.
The long AS--prepending in the first update shows that the advertised AS--path
is a backup path not used at converged state. We explore the backup paths revealed
during the path exploration phenomenon and discover 61.5\% more AS--links not present 
in BTDs.

\begin{table}
  \caption{Example of a simple BGP--update sequence that unveils a backup 
	AS--link (2828 14815) not seen otherwise.}
  \label{example}
  \begin{center}
    \begin{tabular}{|c|c|c|}
      \hline
	Time & AS--path & Prefix \\
      \hline
      \hline
	2003-09-20 12:13:25 &  (withdrawal) & 205.162.1/24 \\
      \hline
	2003-09-20 12:13:55 &  10876-1239-2828-14815-14815-14815-14815-14815 & 205.162.1/24 \\
      \hline
	2003-09-20 12:21:50 &  10876-1239-14815 & 205.162.1/24 \\
      \hline
    \end{tabular}
  \end{center}
\end{table}	

\section{Methodology}
\begin{figure}
\centering
\includegraphics[width=3.5in]{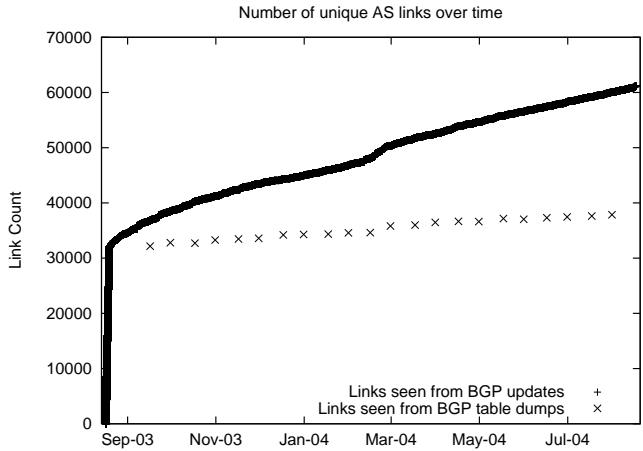}
\caption{Number of unique AS--links observed in BGP updates vs BTDs.}
\label{first_fig}
\end{figure}

Our dataset is comprised of BGP updates collected between September 2003 
and August 2004 from the RV router \texttt{route-views2.oregon-ix.net}. 
The RV router has multihop BGP sessions with 44 BGP routers and saves 
all received updates in the MRT format~\cite{routeviews}.
After converting the updates to ASCII format, we parse the set of 
AS--paths and mark the time each AS--link was first observed, ignoring AS--sets 
and private AS numbers. There are more than 875 million announcements and 
withdrawals, which yield an AS--graph, denoted as $G_{12}$, of 61,134 
AS--links and 19,836 nodes. Subscript 12 in the notation $G_{12}$ refers to the 
number of months in the accumulation period. To quantify the extent of additional 
information gathered from updates, we collect BTDs from the same RV router on the 
1st and 15th of each month between September 2003 and August 2004. For each 
BTD we count the number of unique AS--links, ignoring AS--sets and private AS--numbers 
for consistency. Figure~\ref{first_fig} illustrates the comparison. 
The solid line plots the cumulative number of unique AS--links over time, 
seen in BGP updates. Interestingly, after an initial super--linear increase, 
the number of additional links grows linearly, much faster than the 
corresponding increase observed from the BTDs. At the end of the observation 
window, BGP updates have accumulated an AS--graph that has 61.5\% more 
links and 10.2\% more nodes than the largest BTD--derived graph $G_{12}^{BTD}$, 
which was collected on 08/15/2004. The notable disparity suggests that the real 
Internet AS topology may be different from what we currently observe from
BTD--derived graphs, and merits further investigation. To gain more insight 
in the new information we analyze the temporal and topological properties 
of the AS--connectivity.

\section{Temporal Analysis of Data}

Identifying temporal properties of the AS--connectivity observed from BGP 
updates is necessary to understand the interplay between the observation 
of AS--links and BGP dynamics. In particular, we want to compare the 
temporal properties of AS--links present in BTDs with AS--links observed 
in BGP updates. To do so, we first introduce the concept of {\em visibility} 
of a link from RV. We say that at any given point in time a link is visible 
if RV has received at least one update announcing the link, and the link has 
not been withdrawn or replaced in a later update for the same prefix. A link 
stops been visible if all the prefix announcements carrying the link have been 
withdrawn or reannounced with new paths that do not use the link. We then
define the following two metrics to measure the temporal properties of AS--links:

\begin{enumerate}
\item {\em Normalized Persistence} (NP) of a link is the cumulative time for which a 
  link was visible in RV, over the time period from the first time the link was
  seen to the end of the measurements.
  
\item {\em Normalized Lifetime} (NL) of a link is the time period from the first time  
  to the last time a link was seen, over the time period from the first time the link
  was seen to the end of the measurements.
\end{enumerate}

\begin{figure*} 
  \centerline{
    \subfigure[Normalized Persistence]{\includegraphics[width=2.5in]{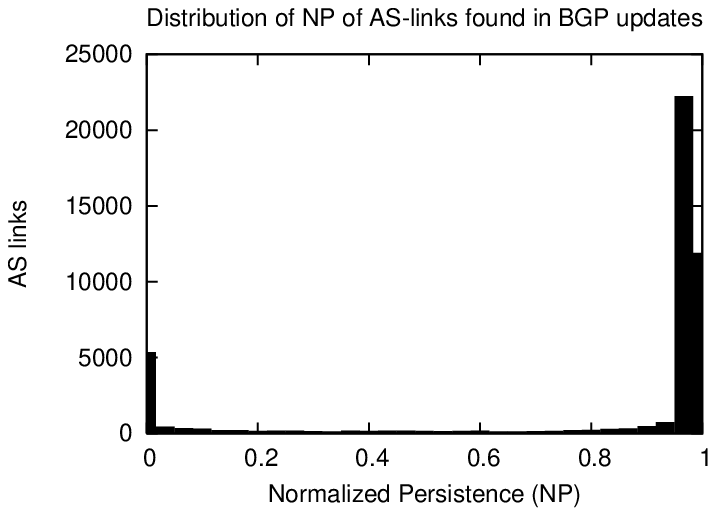}
      \label{NP_figure}}
    \hfil
    \subfigure[Normalized Lifetime]{\includegraphics[width=2.5in]{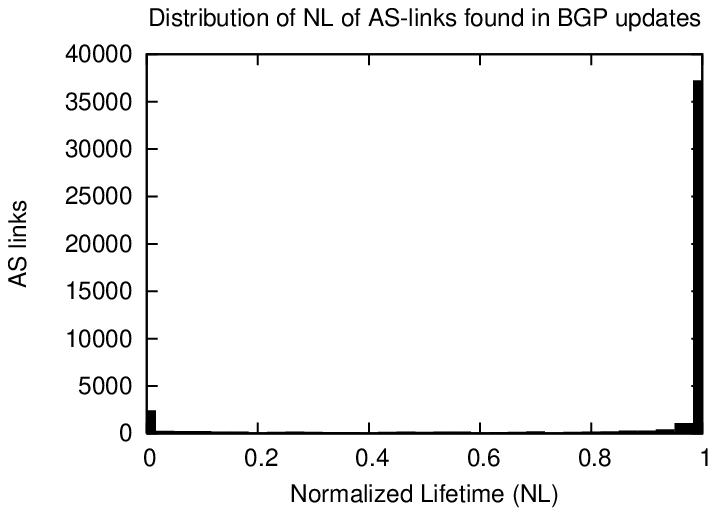}
      \label{NL_figure}}}
  \caption{Distribution of Normalized Persistence and Normilized Lifetime 
	of AS--links seen between September 2003 and January 2004 in BGP updates.}
  \label{second_figure}
\end{figure*}

The NP statistic represents the cumulative time for which a link 
was visible in RV, while the NL represents the span from the beginning 
to the end of the lifetime of the link. Both are normalized over the 
time period from the first time a link was seen to the end of the 
measurements to eliminate bias against links that were not seen 
from the beginning of the observation.

To calculate the NP and NL statistics, we replicate the dynamics of the RV routing
table using the BGP updates dataset. We implement a simple BGP routing daemon that 
parses BGP updates and reconstructs the BGP routing table keeping per--peer and 
per--prefix state as needed. Then for each link we create an array of time 
intervals for which 
the link was visible and calculate the NP and NL statistics. Unfortunately, 
the BGP updates cannot explicitly pinpoint the event of a session reset between RV and
its immediate neighbors. Detection of session resets is necessary to flush invalid
routing table entries learned from the neighbor and to adjust the NP and NL statistics.
We implement a detection algorithm, described in the Appendix, to address the problem.

We measure the NP and NL statistics over a 5--month period, from 
September 2003 to January 2004, and plot their distributions in 
Figure~\ref{second_figure}. Figure~\ref{NP_figure} demonstrates that 
NP identifies two strong modes in the visibility of AS--links.
At the lower end of the $x$ axis, more than 5,000 thousand links 
have $NP \leq 0.2$, portraying that there is a significant number 
of links that only appear during BGP convergence turbulence. At the 
upper end of the $x$ axis, almost 35,000 links have an NP close to 1. 
The distribution~\ref{NL_figure} of the NL statistic is even more modal, 
conveying that most of the links have a high lifetime span. At the end of 
the 5--month period, BGP updates have accumulated a graph $G_{5}$ that we decompose 
into two parts. One subgraph, $G_{5}^{BTD}$, is the topology seen in a BTD collected 
from RV at the end of the 5--month period and the second subgraph is the remaining 
$G_{5} - G_{5}^{BTD}$. Table~\ref{table} shows the number of links with $NP \leq 0.2$, 
$0.2 < NP < 0.8$ and $NP \geq 0.8$ in $G_{5}^{BTD}$ and in $G_{5} - G_{5}^{BTD}$. 
Indeed, only 0.2\% of the links in $G_{5}^{BTD}$ have $NP \leq 0.2$, 
demonstrating that BTDs capture only the AS--connectivity seen at steady--state. 
In contrast, most links in $G_{5} - G_{5}^{BTD}$ have $NP \leq 0.2$, exhibiting 
that most additional links found with our methodology appear during BGP turbulence. 

\begin{table}
  \renewcommand{\arraystretch}{1.3}
  \caption{Normalized Persistence in $G_{5}^{BTD}$ and $G_{5} - G_{5}^{BTD}$.}
  \label{table}
  \begin{center}
    \begin{tabular}{|c||c|c|}
      \hline
      &  $G_{5}^{BTD}$  &  $G_{5} - G_{5}^{BTD}$ \\
      \hline
      \hline
      $NP \leq 0.2$     & 65 (0.2\%) & 6891 (57.5\%) \\
      \hline
      $0.2 < NP < 0.8$  & 1096 (3.2\%) & 1975 (16.5\%) \\ 
      \hline
      $NP \geq 0.8$     &  33141 (96.6\%) & 3119 (26.0\%) \\
      \hline
    \end{tabular}
  \end{center}
\end{table}

\section{Topological Analysis of Data}
Ultimately, we want to know how the new graph is different from the BTD 
graphs, e.g. where the new links are located, and how the properties of 
the graph change. A handful of graph theoretic metrics have been used to 
evaluate the topological properties of the Internet. We choose to evaluate 
three representative metrics of important properties of the Internet topology: 

\begin{enumerate}
      \item {\em Degree Distribution of AS--nodes}. The Internet graph has been 
	shown to belong in the class of power--law networks~\cite{FFF99}. This 
	property conveys the organization principle that few nodes are highly 
	connected. 
      \item {\em Degree--degree distribution of AS--links}. The degree--degree 
	distribution of the AS--links is another structural metric that describes 
	the placement of the links in the graph with respect to the degree of the 
	nodes. More specifically, it is the joint distribution of the degrees of 
	the adjacent ASs of the AS--links.
      \item {\em Betweenness distribution of AS--links}. The betweenness of the 
	AS--links describes the communication importance of the AS--links in the 
	graph. More specifically, it is proportional to the number of shortest 
	paths going through a link.
\end{enumerate}

\begin{figure}
\centering
\includegraphics[width=3.5in]{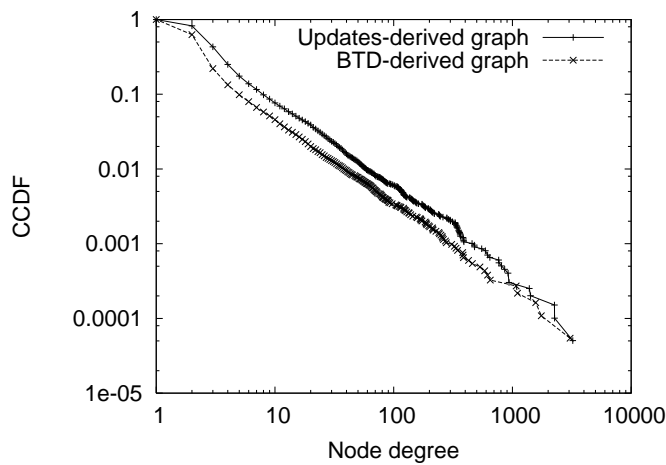}
\caption{CCDF of the AS--degree for the updates--derived AS--graph 
	($G_{12}$) and the largest BTD--derived AS--graph ($G_{12}^{BTD}$).}
\label{ccdf_comp}
\end{figure}

\begin{figure}
\centering
\includegraphics[width=3.5in]{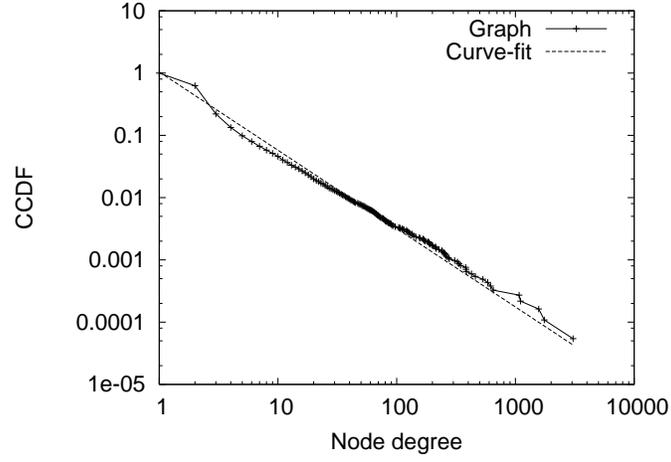}
\caption{CCDF of the AS--degree for the largest BTD--derived 
	AS--graph ($G_{12}^{BTD}$) and linear regression fitted line.}
\label{ccdf_rib}
\end{figure}

\begin{figure}
\centering
\includegraphics[width=3.5in]{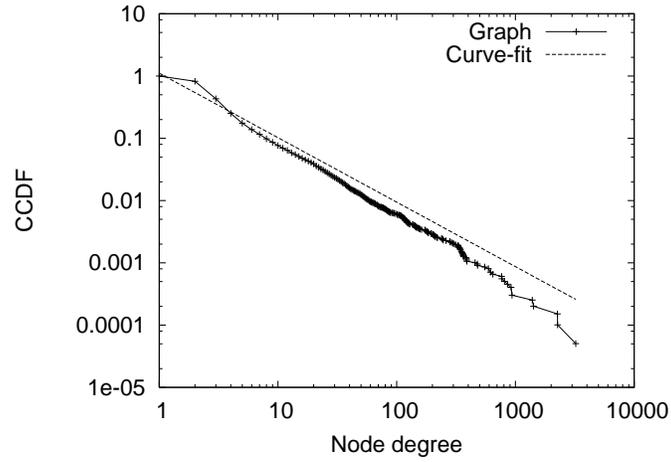}
\caption{CCDF of the AS--degree for the updates--derived AS--graph 
	($G_{12}$) and linear regression fitted line.}
\label{ccdf_updates}
\end{figure}

One of the controversial properties of the Internet topology is that the degree 
distribution of the AS--graph follows a simple power--law expression. This observation 
was first made in~\cite{FFF99} using a BTD--derived AS--graph, later disputed 
in~\cite{chen-origin} using a more complete topology, and finally reasserted 
in~\cite{SFFF03} using an augmented topology as well. Since our work discovers 
substantial additional connectivity over the previous approaches, we re--examine 
the power--law form of the AS--degree distribution. For a power--low distribution 
the complementary cumulative distribution function (CCDF) of the AS--degree is 
linear. Thus, after plotting the CCDF, we can use linear regression to fit a line, 
and calculate the correlation coefficient to evaluate the quality of the fit. 
Figure~\ref{ccdf_comp} plots the CCDF of the AS--degree for the updates-derived 
graph, $G_{12}$, and for the corresponding BTD-derived graph, $G_{12}^{BTD}$.
Due to the additional connectivity in $G_{12}$, the updates--derived curve is 
slightly shifted to the right of the $G_{12}^{BTD}$ curve, without substantial 
change in the shape. Figures~\ref{ccdf_rib} and~\ref{ccdf_updates} show the CCDF 
of the AS--degree and the corresponding fitted line for $G_{12}$ and $G_{12}^{BTD}$, 
accordingly. The correlation coefficient for $G_{12}^{BTD}$ is 0.9836, and in the 
more complete AS--graph $G_{12}$ it slightly decreases to 0.9722, which demonstrates 
that the AS--degree distribution in our updates--derived graph follows a power--law 
expression fairly accurately.

\begin{figure*} 
  \centerline{
    \subfigure[$G_{12}$]{\includegraphics[width=2.5in]{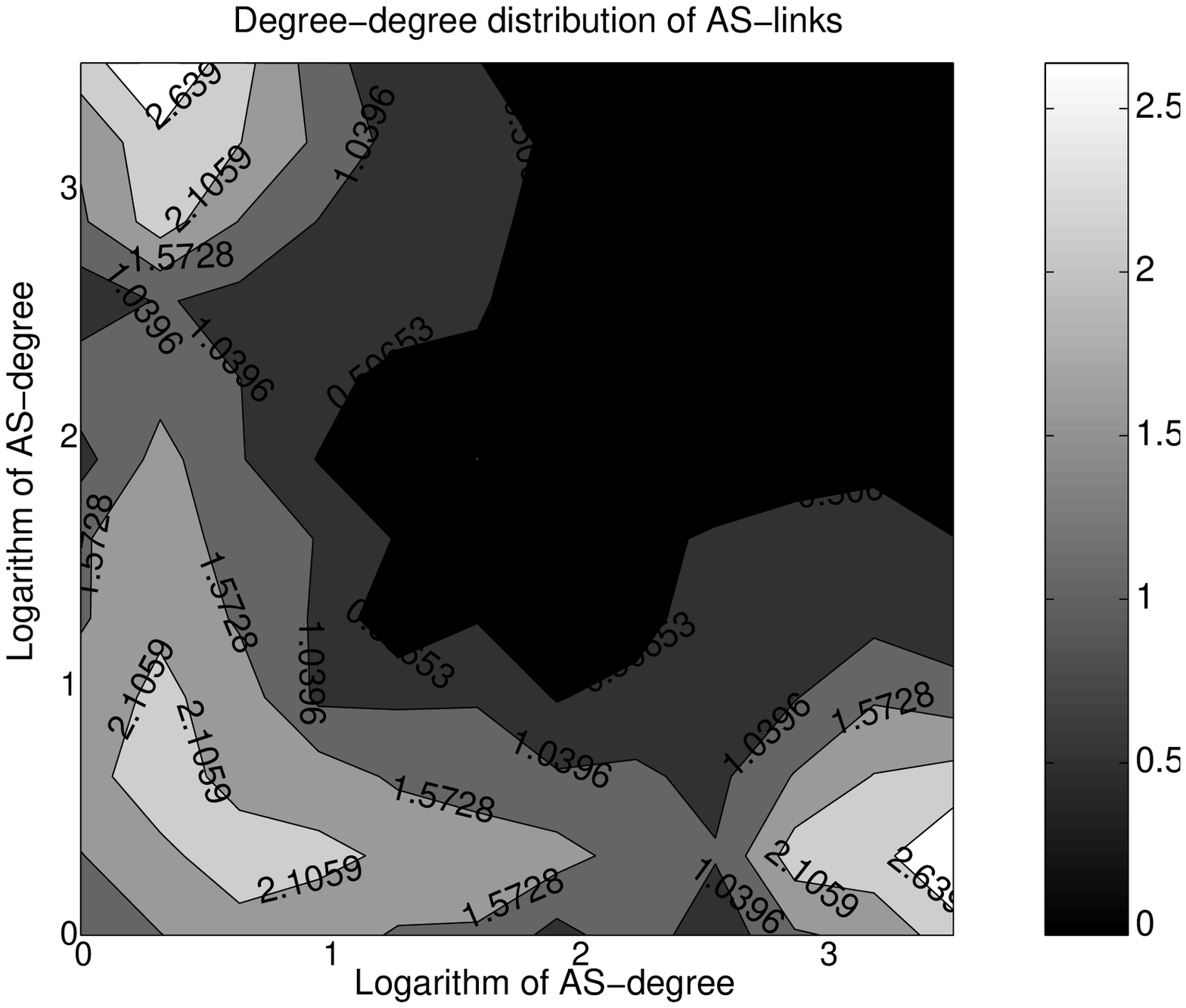}
      \label{full_LC_figure}}
    \hfil
    \subfigure[$G_{12}-G_{12}^{BTD}$]{\includegraphics[width=2.5in]{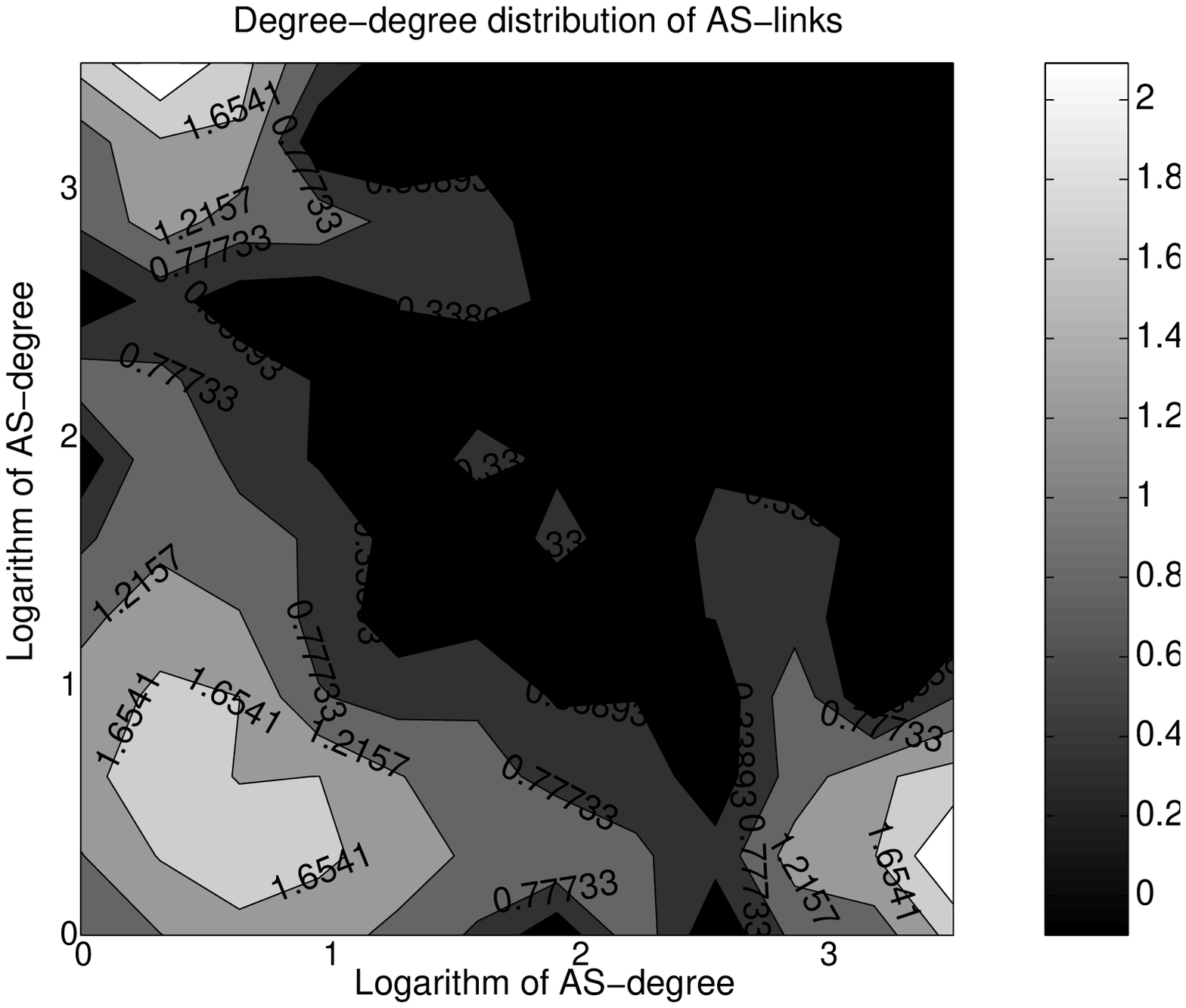}
      \label{diff_LC_figure}}}
  \caption{Degree--degree distributions of AS--links. The $x$ and $y$ axes show 
	the logarithms of the degrees of the nodes adjacent to a link. The color 
	codes show the logarithm of the number of the links connecting ASs of 
	corresponding degrees.}
  \label{degree_degree_figure}
\end{figure*}

\begin{figure}
\centering
\includegraphics[width=3.5in]{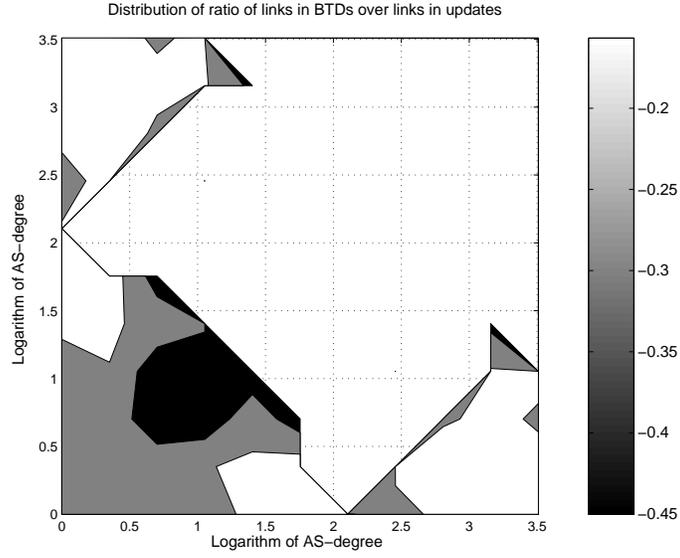}
\caption{Distribution of the ratio of the number of links in ~$G_{12}^{BTD}$ over
	the number of links in~$G_{12}$ connecting ASs of corresponding degrees.
	The $x$ and $y$ axes show the logarithms of the degrees of the nodes 
	adjacent to a link. The color codes show the logarithm of the above ratio.}
\label{T2UR}
\end{figure}

We then examine the degree--degree distribution of the links. The degree--degree 
distribution~$M(k_1,k_2)$ is the number of links connecting ASs of degrees 
$k_1$ and $k_2$. Figure~\ref{degree_degree_figure}, compares the
degree--degree distributions of the links in the full~$G_{12}$ graph and of the links
present only in updates, $G_{12} - G_{12}^{BTD}$. The overall structure of the two
contourplots is similar, except for the differences in the areas of links connecting
low-degree nodes to low-degree nodes and links connecting medium-degree nodes to
medium-degree nodes (the bottom-left corner and the center of the contourplots).
The absolute number of such links in~$G_{12} - G_{12}^{BTD}$ is smaller than 
in~$G_{12}$, since~$G_{12} - G_{12}^{BTD}$ is a subgraph of ~$G_{12}$. However, 
the contours illustrate that the ratio of such links in~$G_{12} - G_{12}^{BTD}$ 
to the total number of links in~$G_{12} - G_{12}^{BTD}$ is higher than the 
corresponding ratio of links in~$G_{12}$.
Figure~\ref{T2UR} depicts the contourplot of the ratio of the number of links
in~$G_{12}^{BTD}$ over the number of links in~$G_{12}$ connecting ASs of 
corresponding degrees. The dark region between 0.5 and 1.5 exponents on 
the $x$ and $y$ axes, signifies the fact that BGP updates contain additional links,
compared to BTDs, between low and medium-degree ASs close to the periphery of 
the graph.

\begin{figure}
\centering
\includegraphics[width=3.5in]{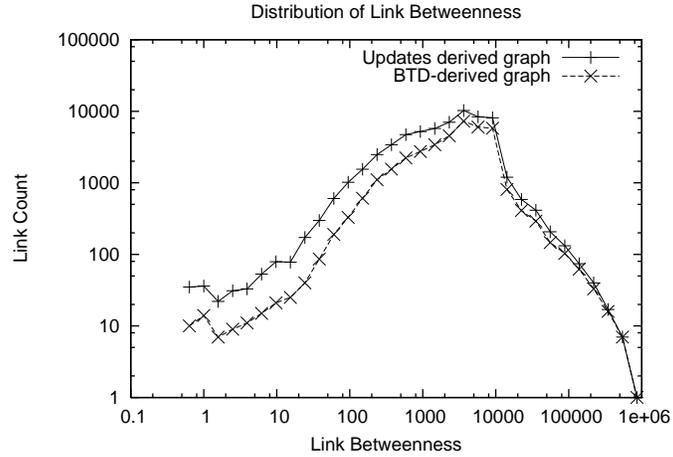}
\caption{Distribution of the link betweenness of $G_{12}$ compared to $G_{12}^{BTD}$.}
\label{BW_dstr}
\end{figure}

Finally, we examine the link betweenness of the AS--links. In graph~$G(V,E)$, 
the betweenness~$B(e)$ of link~$e \in E$ is defined as

\begin{equation}
B(e) = \sum_{ij \in V}\frac{\sigma_{ij}(e)}{\sigma_{ij}},\nonumber 
\end{equation}
where~$\sigma_{ij}(e)$ is the number of shortest paths between nodes~$i$ and~$j$ 
going through link~$e$ and~$\sigma_{ij}$ is the total number of shortest paths 
between~$i$ and~$j$. With this definition, link betweenness is proportional to 
the traffic load on a given link under the assumptions of uniform traffic 
distribution and shortest--path routing. Figure~\ref{BW_dstr} illustrates 
the betweenness distribution of $G_{12}$ and of $G_{12}^{BTD}$ and reveals 
that our updates--constructed graph yields more links with small betweenness. 
Links with small betweenness have lower communication importance in a graph 
theoretic context, demonstrating that our methodology unveils backup links 
and links used for local communication in the periphery of the graph.

Overall, our topological analysis shows that our augmented graph remains a power-law
network and has more links between low and medium--degree nodes and more links of
lower communication importance compared to BTD--derived graphs.

\section{Conclusions}

In this work we exploit the previously unharnessed topological information 
that can be extracted from the most well--known and easily accessible source 
of Internet interdomain routing data. We evidence that the Internet topology 
is vastly larger than the common BTD--derived topologies and we show how an 
undesired aspect of the interdomain architecture can be used constructively. 
We find that our substantially larger AS--graph retains the power--law property 
of the degree distribution. Finally, we show that our method discovers links 
of small communication importance connecting low and medium--degree ASs, 
suggesting AS--links used for backup purposes and local communication in the 
periphery of the Internet.

Closing, we highlight that our work is a step forward showing a large 
gap in our knowledge of the Internet topology. For this reason, we 
pronounce the need to focus more on the perpetual problem of measuring 
Internet topology before accepting far--reaching conclusions based on 
currently available AS--level topology data, which are undeniable rich 
but substantially incomplete.

\section*{Acknowledgments}

We thank Priya Mahadevan for sharing her betweenness scripts, Andre Broido, 
Bradley Huffaker and Young Hyun for valuable suggestions, and Spyros Denazis 
for providing computer resources.

Support for this work was provided by the DARPA N66002-00-1-8934, NSF award 
number CNS-0427700 and CNS-0434996.


\section*{APPENDIX}
\label{APPENDIX}

\subsection*{Detection of session resets}
The problem of detection of BGP session resets has also been addressed by others.
In~\cite{MaFe02} Maennel {\em et al.} propose a heuristic to detect session resets 
on AS--links in arbitrary Internet locations by monitoring BGP updates in RV. We 
are concerned with a seemingly less demanding task: detection of session resets 
with immediate neighbors of RV. Our algorithm is composed of two components. The 
first detects surges in the BGP updates received from the same peer over a short 
time window of $s$ seconds. If the number of unique prefixes updated in $s$ are 
more than a significant percent $p$ of the previously known unique prefixes from 
the same peer, then a session reset is inferred. The second component detects 
periods of significant inactivity when a threshold $t$ is passed from otherwise 
active peers. We combine both approaches and set low thresholds ($t = 4mins$, 
$p = 80\%$, $s = 4 secs$) to yield an aggressive session reset detection algorithm. 
Then, we calculate NP and NL over a period of a month with and without aggressive 
session reset detection enabled. We find that the calculated statistics are 
virtually the same with less then 0.1\% variation. Implying that the short time 
scale of the lifetime of session resets does not affect the span of the NP and 
NL statistics. Hence, we leave out the detection of session resets in the remaining 
NP and NL measurements. 
\end{document}